\documentclass[fleqn,usenatbib,useAMS]{mnras}

\usepackage{graphicx}	
\usepackage{amsmath}	
\usepackage{amssymb}	
\usepackage{multicol}        
\usepackage{bm}		
\usepackage{pdflscape}	

\usepackage[T1]{fontenc}
\usepackage{ae,aecompl}

\usepackage{newtxtext,newtxmath}

\title[Variable Dust Emission from GD\,56]{Dust Production and Depletion in Evolved Planetary Systems}

\author[J. Farihi et al.]{J. Farihi$^1$\thanks{E-mail: j.farihi@ucl.ac.uk},
R. van Lieshout$^2$,
P. W. Cauley$^{3,11}$,
E. Dennihy$^4$,
K. Y. L. Su$^5$,
S. J. Kenyon$^6$,
\newauthor
T. G. Wilson$^{1,7}$,
O. Toloza$^8$,
B. T. G\"ansicke$^{8,9}$,
T. von Hippel$^{10}$,
S. Redfield$^{11}$,
J. H. Debes$^{12}$,
\newauthor
S. Xu$^{13}$,
L. Rogers$^2$,
A. Bonsor$^2$,
A. Swan$^1$,
A. F. Pala$^8$,
W. T. Reach$^{13}$
\medskip
\\
$^1$Department of Physics and Astronomy, University College London, London WC1E 6BT, UK\\
$^2$Institute of Astronomy, University of Cambridge, Cambridge CB3 0HA, UK\\
$^3$School of Earth and Space Exploration, Arizona State University, Tempe, AZ 85287, USA\\
$^4$Physics and Astronomy Department, University of North Carolina at Chapel Hill, Chapel Hill, NC 27599, USA\\
$^5$Steward Observatory, University of Arizona, Tucson AZ 85721, USA\\
$^6$Smithsonian Astrophysical Observatory, Cambridge, MA 02138, USA\\
$^7$Isaac Newton Group of Telescopes, E-38700 Santa Cruz de La Palma, Spain\\
$^8$Department of Physics, University of Warwick, Coventry CV4 7AL, UK\\
$^9$Centre for Exoplanets and Habitability, University of Warwick, Coventry CV4 7AL, UK\\
$^{10}$Physical Sciences Department, Embry-Riddle Aeronautical University, Daytona Beach FL 32114, USA\\
$^{11}$Astronomy Department and Van Vleck Observatory, Wesleyan University, Middletown 06459, USA\\
$^{12}$Space Telescope Science Institute, Baltimore, MD 21218, USA\\
$^{13}$Gemini Observatory, Northern Operations, Hilo, HI 96720, USA\\
$^{14}$SOFIA-USRA, NASA Ames Research Center, Moffett Field, CA 94035, USA
}

\begin{document}


\maketitle

\begin{abstract}
The infrared dust emission from the white dwarf GD\,56 is found to rise and fall by 20\% peak-to-peak over 11.2\,yr, and is 
consistent with ongoing dust production and depletion.  It is hypothesized that the dust is produced via collisions associated 
with an evolving dust disk, temporarily increasing the emitting surface of warm debris, and is subsequently destroyed or 
assimilated within a few years.  The variations are consistent with debris that does not change temperature, indicating that
dust is produced and depleted within a fixed range of orbital radii.  Gas produced in collisions may rapidly re-condense onto 
grains, or may accrete onto the white dwarf surface on viscous timescales that are considerably longer than Poynting-Robertson 
drag for micron-sized dust.  This potential delay in mass accretion rate change is consistent with multi-epoch spectra of the 
unchanging Ca\,{\sc ii} and Mg\,{\sc ii} absorption features in GD\,56 over 15\,yr, although the sampling is sparse.  Overall
these results indicate that collisions are likely to be the source of dust and gas, either inferred or observed, orbiting most or 
all polluted white dwarfs.
\end{abstract}


\section{Introduction}

Observable and real-time changes in exoplanetary systems hold important clues for dynamical processes during their birth 
and long-term evolution, where the number of variable systems is likely to increase owing to large ground- and space-based 
surveys.  Sensitive monitoring of giant planetary and substellar atmospheres can reveal periodic features such as rotation, 
atmospheric wind speeds, and global weather patterns \citep{sne14,lou15,apa17}.  Smaller major and minor exoplanetary
bodies are typically out of reach for real-time monitoring, but those that actively produce debris can generate sufficient area
to be detected via absorption or emission and studied over time.  Such systems have the potential to constrain the bigger
picture of planet formation and evolution, especially if they exhibit secular changes, non-periodic events, or signposts of 
important evolutionary phases such as the Late Heavy Bombardment.

Active exo-cometary populations are known in a handful of systems via transient absorption in optical and ultraviolet spectra 
\citep{kie14a,kie14b,wel18}, while {\em Kepler} data now includes convincing transits of individual exo-comets in at least one 
system \citep{rap18}.  The dramatic and irregular flux changes measured towards KIC\,8462852 are also broadly consistent 
with an exo-cometary origin \citep{boy18,wya18}.  The inner and terrestrial planet-analog regions are more challenging to 
detect in general due to their host star proximity, but sensitivity to compact orbits has enabled {\em Kepler} to detect transits 
from a trio of rocky planets via tailing debris clouds associated with their short periods and thus irradiation-driven mass loss 
\citep{van17}.  Analogous processes may contribute to the transit light curves of the polluted white dwarf WD\,1145+017 
(hereafter WD\,1145; \citealt{van15,gan16,rap16}).

Exoplanetary system variability has also been observed in light curves via emission, including a few spectacular examples 
of infrared flux changes associated with the terrestrial-planet forming regions \citep{mel12,men14}.  And within the former
terrestrial zones of A-type and similar stars, there is myriad evidence for rocky planetesimal activity via the evolved planetary 
systems orbiting and polluting white dwarf stars (see \citealt{far16} and references therein).  The first clear evidence of any
variability in these systems emerged via changes in disk line emission \citep{gan08}, and preceded transit detections by 
several years.  Similar changes have now been documented for five disks via gas emission or absorption \citep{wil14,
man16a,man16b,den18}, including WD\,1145 \citep{red17,cau18}, but to date only a single white dwarf system has been 
shown to vary in the infrared.

SDSS\,J095904.69-020047.6 (hereafter SDSS\,0959) was reported to show a drop in $3-5\,\upmu$m infrared flux by 
around 35\% in less than 300\,d \citep{xu14} based on the comparison of its warm {\em Spitzer} IRAC discovery fluxes 
\citep{far12a} to measurements made with {\em WISE}.  \citet{xu14} surmised the drop in flux was likely due to a change 
in the inner disk radius from an impact or instability, causing a few percent of the entire disk mass to be accreted within 
the duration of the observed change, and suggest that the resulting accretion rate increase could lead to an observable 
difference in the photospheric metal abundance on similarly-short timescales.

This paper reports long-term, $3-5\,\upmu$m infrared flux variations in the polluted white dwarf GD\,56 (= WD\,0408--041),
a hydrogen atmosphere (DA-type) star with $T_{\rm eff}\approx15\,000$\,K and a cooling age around 200\,Myr \citep{gia11}.
The variability in the infrared is presented together with optical spectroscopy that reveals constant, photospheric metal 
absorption over a similar timescale.  Infrared data are comprised of nine {\em Spitzer} observational epochs spanning 
11.2\,yr from 2006 to 2017, and are supplemented by multi-epoch {\em WISE} data from 2010 and 2014--2017.  There 
has been one substantial increase of at least 20\%, and what appear to be two decaying trends of similar magnitude, 
all taking place over several years each.  These changes are interpreted as the production (increase) of dust clouds, and 
their subsequent depletion (decrease), where possible scenarios may also account for the flux decrease at SDSS\,0959.  

The rest of the paper is organized as follows.  Section two describes the infrared data, where the analysis includes both 
absolute flux measurements and differential photometry using field stars, and the spectroscopic data that indicate no 
appreciable changes to the inferred metal accretion rate.  Section three describes the (weak) constraints on disk models 
based on the data, as well as theoretical considerations, and discusses possible model families for GD\,56 and dusty 
white dwarfs in general.  Section four gives the summary and outlook.

\section{OBSERVATIONS AND ANALYSIS}

The following section describes the two types of observational data for GD\,56 used in this study.  The first are infrared 
photometric observations of excess emission that track the properties of warm circumstellar dust, and include data taken 
with the {\em Spitzer Space Telescope} \citep{wer04}, the {\em Wide-field Infrared Survey Explorer (WISE}; \citealt{wri10}),
and ground-based $JHK$ photometry.  The second are medium- to high-resolution optical spectra that trace the abundance 
of atmospheric metals that result from disk accretion.  The infrared and optical data sets were independently obtained by 
several teams for various purposes; there is no correlated timing between any.  In particular, the {\em Spitzer} and optical 
data are both irregularly distributed in time, but the NEOWISE data have semi-regular cadence.

\subsection{Infrared Photometry}

Multi-epoch {\em Spitzer} IRAC \citep{faz04} data were retrieved from the archive, where they were first processed by 
the {\em Spitzer} Science Center via pipeline S18.18.0 for cryogenic data, or S19.2.0 for warm mission data.  There are 
six programs\footnote{ID\# 275, 40369, 90095, 10032, 10175, 13216} that targeted GD\,56 between 2006 Sep and 2017 
Dec, and that yield nine total epochs of data in either channel 1 at 3.6\,$\upmu$m or channel 2 at 4.5\,$\upmu$m.  This
study focuses on the two shortest wavelength bandpass data, as the four most recent epochs were taken post-cryogen, 
and moreover they are directly comparable to channels 1 and 2 of {\em WISE}.  Together these collective data yield a 
compelling data set.

\begin{figure}
\includegraphics[width=84mm]{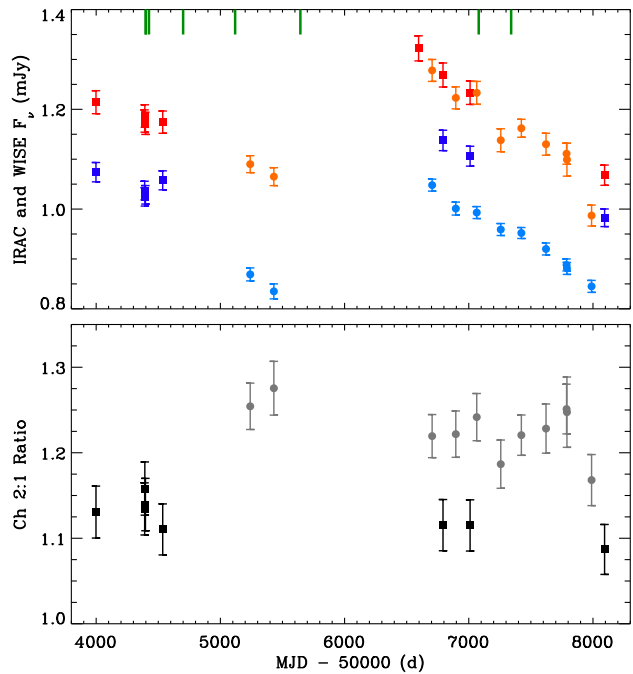}
\caption{Flux-calibrated infrared data for GD\,56 over 11.2\,yr as measured by {\em Spitzer} and {\em WISE} in their 
two shortest wavelength channels.  In the upper panel, the IRAC and {\em WISE} fluxes with errors are given as squares 
and circles, respectively, with blue hues indicating channel 1 centered near 3.5\,$\upmu$m, and red hues for channel 
2 centered near 4.5\,$\upmu$m.  The corresponding pairs of bandpass filters on each spacecraft are similar, but vary 
by a few percent in isophotal wavelength and by several percent in bandwidth \citep{jar11}.  All IRAC photometry has 
S/N $\gg100$ so that the measurements are limited by the $\approx2$\% calibration uncertainty of the instrument.  
Each {\em WISE} flux is the weighted mean and error of multi-epoch photometry in AllWISE and NEOWISE, where 
each epoch typically consists of 12 individual measurements.  The green vertical bars at the top of the panel indicate 
eight dates where spectroscopic observations of GD\,56 were taken.  The lower panel plots the channel 2:1 flux ratios, 
with IRAC data in black and the {\em WISE} ratios in grey.  These ratios indicate that the temperature of the emitting 
dust -- and hence its location -- does not appear to be changing significantly.
\label{fig1}}
\end{figure}

Fluxes at 3.6 and 4.5\,$\upmu$m were each measured in two ways; first on single exposure, basic calibrated data (BCD) 
frames, and second on mosaics created using {\sc mopex} following the best practices as detailed in the IRAC Instrument 
Handbook v2.1.  Both {\sc iraf} and {\sc apex} were used to perform aperture photometry on GD\,56, where photometry 
was executed using aperture radii of 2--3 native pixels and sky annuli of 12--20 native pixels with appropriate aperture 
corrections.  The signal-to-noise ratios (S/N) of these data are on the order of several hundred or greater and thus 
contribute negligible measurement error to the flux uncertainties, and thus the total errors are limited by the calibration 
uncertainty of the instrument ($\approx2$\%; \citealt{rea05}).  As these results only depend on the changing flux for a 
source within a given passband, the calibration uncertainty is irrelevant.  The different methods of measuring flux via 
aperture photometry were all found to agree to $\ll1$\%, and the adopted values are plotted in Figure \ref{fig1} and 
listed in Table \ref{tbl1}.

\begin{figure}
\includegraphics[width=84mm]{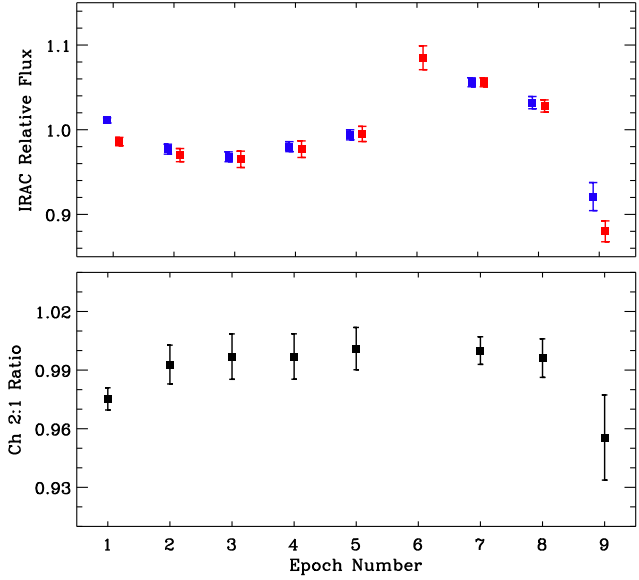}
\caption{Relative IRAC fluxes obtained via differential photometry for GD\,56 in both channels using five comparison 
stars in common for all nine {\em Spitzer} epochs.  Symbols and colors are the same as in Figure \ref{fig1}, and there
are no data available in IRAC channel 1 for the sixth epoch (program ID 90095).  The upper panel plots the weighted 
mean and error of five differential photometric measurements per epoch, where the values are normalized so that the 
average of all epochs is 1.0 in each channel, and with small horizontal offsets between the 3.6 and 4.5\,$\upmu$m 
data points for clarity.  These relative fluxes were measured for all sources identically as described in Section 2.1, 
and include only errors propagated based on measurement uncertainties (i.e.\ S/N).  Any intrinsic variation among 
the comparison stars is not accounted for here, but ignoring this possibility yields a difference of 23\% between the 
minimum and maximum differential measurements, at over $10\,\upsigma$.  The lower panel plots the ratios of the 
channel 2:1 differential photometry from the upper panel, and again suggests the temperature of the dust is not 
changing over the long timescale probed by these widely-spaced observations.
\label{fig2}}
\end{figure}

To better constrain the significance of the actual changes, the IRAC image mosaics were used to perform differential 
photometry on GD\,56.  The brightest five, isolated field stars with good overlap (mosaic) coverage were chosen for 
relative flux measurements, and are all around 2--5 times fainter than the science target in both channels.  Flux ratios 
between each field star and the science target were determined using {\sc apex} aperture photometry as above with 
$r=2$ native pixels.  The uncertainties of individual relative fluxes were calculated using the quadrature sum of the 
measurement errors for each pair.  For each IRAC epoch the weighted mean and error of these five ratios are plotted
in Figure \ref{fig2}, after normalization across all epochs.  While these relative fluxes appear to strongly confirm the real 
nature of the flux changes in GD\,56, the measurement errors alone may underestimate the true uncertainty as these 
do not account for potential bona fide variations in the comparison field stars.

\begin{table}
\begin{center}
\caption{Infrared Observations of GD\,56\label{tbl1}}
\begin{tabular}{@{}cccc@{}}

\hline

	&\multicolumn{3}{c}{{\em Spitzer} IRAC}\\

Date			&Ch\,1: 3.55\,$\upmu$m	&Ch\,2: 4.49\,$\upmu$m	&Ch\,2 / Ch\,1\\
			&(mJy)					&(mJy)					&\\
\hline			
			
2006.09.20	&$1.074\pm0.019$ 			&$1.214\pm0.023$			&$1.130\pm0.030$\\
2007.10.17	&$1.037\pm0.019$ 			&$1.177\pm0.022$			&$1.134\pm0.030$\\
2007.10.18	&$1.025\pm0.018$			&$1.187\pm0.023$			&$1.158\pm0.031$\\
2007.10.23	&$1.029\pm0.019$ 			&$1.172\pm0.022$			&$1.139\pm0.031$\\
2008.03.10	&$1.058\pm0.019$  			&$1.174\pm0.022$			&$1.110\pm0.030$\\
2013.10.30	&...						&$1.322\pm0.025$			&...\\
2014.05.15	&$1.138\pm0.020$ 			&$1.269\pm0.024$			&$1.115\pm0.030$\\
2014.12.19	&$1.106\pm0.020$ 			&$1.233\pm0.023$			&$1.115\pm0.030$\\
2017.12.06	&$0.983\pm0.018$			&$1.068\pm0.020$			&$1.087\pm0.029$\\

\\
	&\multicolumn{3}{c}{{\em WISE} + NEOWISE}\\

			&Ch\,1: 3.35\,$\upmu$m	&Ch\,2: 4.60\,$\upmu$m	&Ch\,2 / Ch\,1\\
			&(mJy)					&(mJy)					&\\
\hline

2010.02.13	&$0.869\pm0.013$			&$1.090\pm0.017$			&$1.254\pm0.027$\\
2010.08.22	&$0.835\pm0.015$			&$1.065\pm0.018$			&$1.275\pm0.031$\\
2014.02.16	&$1.048\pm0.012$			&$1.278\pm0.022$			&$1.219\pm0.025$\\
2014.08.26	&$1.001\pm0.013$			&$1.223\pm0.022$			&$1.222\pm0.027$\\
2015.02.11	&$0.993\pm0.012$			&$1.233\pm0.023$			&$1.242\pm0.028$\\
2015.08.23	&$0.959\pm0.012$			&$1.138\pm0.023$			&$1.187\pm0.028$\\
2016.02.05	&$0.952\pm0.011$			&$1.162\pm0.018$			&$1.220\pm0.024$\\
2016.08.21	&$0.920\pm0.012$			&$1.130\pm0.022$			&$1.228\pm0.029$\\
2017.02.03	&$0.888\pm0.012$			&$1.111\pm0.021$			&$1.251\pm0.029$\\
2017.02.06	&$0.881\pm0.012$			&$1.099\pm0.033$			&$1.247\pm0.041$\\
2017.08.23	&$0.845\pm0.012$			&$0.987\pm0.021$			&$1.168\pm0.030$\\

\\
	&\multicolumn{3}{c}{Ground-Based Photometry}\\

					&$J$				&$H$			&$K$\\ 
					&(mag)           		&(mag)			&(mag)\\ 
\hline

1998.10.12$^{\rm a}$	&$15.87\pm0.06$ 	&$15.99\pm0.13$	&$15.44\pm0.18$\\	
2006.10.09$^{\rm b}$	&$15.85\pm0.05$	&$15.75\pm0.05$	&$15.13\pm0.05$\\
2014.10.03$^{\rm c}$	&$15.96\pm0.03$	&$15.76\pm0.04$	&$15.10\pm0.04$\\
2016.10.10$^{\rm c}$  	&$15.95\pm0.03$	&$15.77\pm0.03$	&$15.12\pm0.03$\\
2017.01.31$^{\rm c}$  	&$15.92\pm0.03$	&$15.80\pm0.04$	&$15.16\pm0.03$\\
   
\hline

\end{tabular}
\end{center}

\flushleft
$^{\rm a}$ 2MASS \citep{skr06}.

$^{\rm b}$ IRTF; SpeX \citep{far09b}.

$^{\rm c}$ UKIRT; WFCAM (this work).

\end{table}

\begin{figure}
\includegraphics[width=84mm]{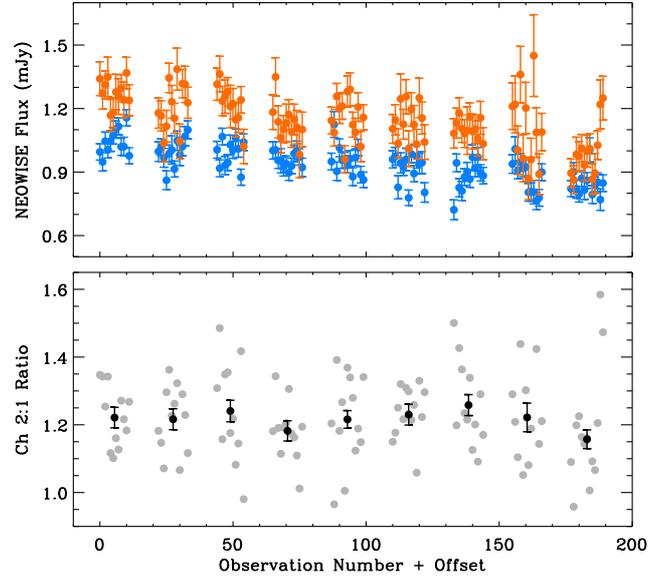}
\caption{All NEOWISE measurements for GD\,56 in channels 1 and 2, where the colors are the same as in Figure \ref{fig1}.
These are 110 individual fluxes and errors, where typically one dozen are obtained during a single epoch spanning many 
hours, and epochs are separated by around 180\,d on average.  The data are displayed in chronological order by number, 
with distinct epochs offset along the x-axis.  The dimming trend is clear in the raw data of both the 3.4 and 4.6\,$\upmu$m 
channels, but the flux ratio in the bottom panel does not show any significant change or trend, despite the regular and
roughly semi-annual sampling of this observational subset.  The black data points with error bars are the weighted mean
and weighted error of the individual flux ratios (and their errors, not shown) plotted in in grey.
\label{fig3}}
\end{figure}

Additional multi-epoch infrared photometry was taken by {\em WISE} and retrieved from the AllWISE and NEOWISE 
source catalogs.  During the cryogenic mission there were two sets of observations for GD\,56 (2010 Feb and Aug), and 
during NEOWISE \citep{mai11} there have been an additional nine, regularly-spaced epochs (2014 Feb to 2017 Aug).  
Each epoch of data consists of approximately one dozen individual measurements and errors, and for each, both the 
fluxes and their corresponding errors were determined by weighted average.  Typical uncertainties in mean {\em WISE} 
flux are comparable to the IRAC data at $\approx2$\%, and combined data points for each epoch are shown in Figure 
\ref{fig1}.  The NEOWISE data are essentially contiguous with a cadence near either 165 or 195\,d, and the individual 
measurements and errors are shown in Figure \ref{fig3}.

All absolute flux measurements are plotted in Figure \ref{fig1} as a function of date, as well as the flux ratios between the
longer and shorter wavelength channels.  Because the two spacecraft have similar but still distinct filter sets, the channel
2:1 flux ratios of IRAC and {\em WISE} cannot be directly compared.  Furthermore, while {\em Spitzer} and {\em WISE} 
share a network of infrared calibration standards, the absolute scales differ by up to a few per cent \citep{jar11}.  The 
$2-20$\,$\upmu$m continuum emission from GD\,56 has previously been shown to be consistent with a $T=1000$\,K 
blackbody \citep{jur07}, and this model flux can be convolved with the filter bandpasses of both spacecraft, providing 
an estimated transformation between {\em WISE} and IRAC fluxes and flux ratios.  Such a transformation can roughly
reproduce both the fluxes and flux ratios where observations occurred most closely together in time.  The best such 
mapping is achieved near a blackbody temperature of 900\,K, where all fluxes and flux ratios between spacecraft are 
then within $1\upsigma$.

Lastly, GD\,56 was observed in the $1-2.5\,\upmu$m region between 2014 Oct and 2017 Jan using the Wide Field 
Camera (WFCAM) on the United Kingdom Infrared Telescope (UKIRT) on Mauna Kea (Rogers et al.\ 2018, in prep.). 
For each of the $JHK$ bands, three epochs of imaging were obtained using a five-point dither pattern and a total 
exposure time of 75\,s.  The infrared images were pipeline processed by the Cambridge Astronomical Survey Unit to 
produce photometrically and astrometrically calibrated images and source catalogs \citep{hod09}.  For each observing 
epoch, the weighted mean $JHK$ magnitudes for GD\,56 were taken from the output catalogs, and to the photometric 
measurement errors was added a systematic uncertainty typical of infrared arrays (e.g.\ \citealt{leg06}).  The resulting 
WFCAM photometry is reported in Table \ref{tbl1}, together with previously published $1-2.5\,\upmu$m data.  All the 
measured $JHK$ photometry are consistent within $2\upsigma-3\upsigma$ in each of the bandpasses, and the 
values themselves generally vary by less than around 6\%.

\subsection{Optical Spectroscopy}

\begin{figure}
\includegraphics[width=84mm]{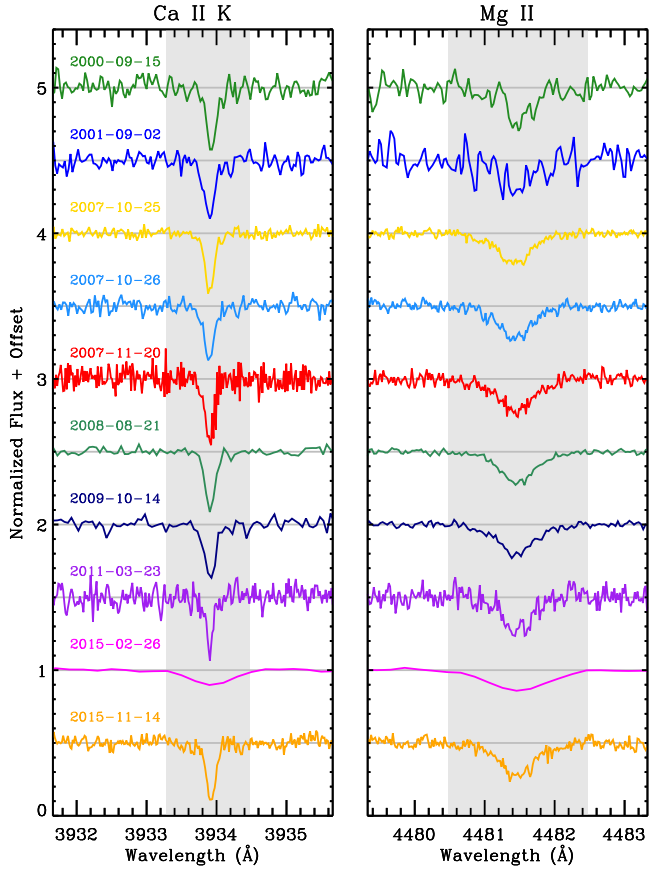}
\caption{Multi-epoch optical spectroscopy of GD\,56 in the region of the Ca\,{\sc ii} K and Mg\,{\sc ii} 4481\,\AA \ lines, 
with details provided in Table \ref{tbl2}.  All spectra are normalized and vertically offset for clarity, with chronological 
ordering from top down.  The 2011 Mar 23 spectrum exhibits weaker and narrower lines, but is the result of a single 
exposure taken during twilight, with bright sky emission throughout the wavelength regions of interest.  Hence, this may 
have affected the fidelity of the data during the spectral extraction process.  The velocities of all absorption features are 
consistent within the wavelength calibration errors.
\label{fig4}}
\end{figure}

Echelle spectroscopy of GD\,56 with sufficient resolution to detect atmospheric metals was executed using five separate 
instruments over ten epochs between the photospheric metal discovery spectrum in 2000 Sep \citep{koe05} and 2015 Nov.  
The relevant observational details are summarized in Table \ref{tbl2}.  Individual exposures were extracted, combined as
appropriate, and calibrated according to the facility-recommended methods and software (e.g.\ \citealt{kel00,kel03,fre13}).  
Spectra were all continuum normalized and portions of each are shown in Figure \ref{fig4}.  The Keck HIRES spectrum 
taken on 2011 Mar 23 is a single exposure that was completed prior to the end of twilight and suffers from high sky counts
-- including emission lines -- in all relevant orders.  For the remaining spectra there were no significant observing or 
instrument conditions that may have affected the data.

\begin{table*}
\begin{center}
\caption{Summary of Blue Optical Spectroscopic Observations and Analyses for GD\,56\label{tbl2}}
\begin{tabular}{@{}lllcrcllcc@{}}

\hline

Date				&Facility		&Instrument	&Slit	Size	&Resolving 	&S/N		&$W_{\rm Ca\,II\,K}$	&$W_{\rm Mg\,II}$	&[Ca/H]$^{\rm a}$	&[Mg/H]$^{\rm a}$\\
				&			&			&(arcsec)	&Power	&($\uplambda=4000$\,\AA)&(m\AA)			&(m\AA)		&				&\\
				
\hline

2000.09.15		&VLT		&UVES		&2.1		&20\,000		&19		&$81\pm13$		&$125\pm19$		&$-6.9\pm0.1$		&$-5.5\pm0.1$\\
2001.09.02		&VLT		&UVES		&2.1		&20\,000		&18		&$81\pm13$		&$126\pm22$		&$-7.0\pm0.1$		&$-5.5\pm0.1$\\
2007.10.25		&Magellan	&MIKE		&0.7		&40\,000		&37		&$72\pm7$		&$133\pm5$		&$-7.1\pm0.2$		&$-5.6\pm0.1$\\
2007.10.26		&Magellan	&MIKE		&0.7		&40\,000		&27		&$75\pm9$		&$137\pm6$		&$-7.2\pm0.1$		&$-5.6\pm0.1$\\  
2007.11.20		&Keck		&HIRES		&1.1		&36\,000		&13		&$74\pm13$		&$152\pm9$		&$-7.1\pm0.2$		&$-5.6\pm0.1$\\    
2008.08.21		&Magellan	&MIKE		&0.7		&40\,000		&44		&$76\pm7$		&$144\pm5$		&$-7.0\pm0.2$		&$-5.6\pm0.1$\\   
2009.10.14		&Magellan	&MIKE		&0.7		&40\,000		&38		&$84\pm11$		&$139\pm6$		&$-7.1\pm0.2$		&$-5.6\pm0.1$\\
2011.03.23$^{\rm b}$&Keck		&HIRES		&1.1		&36\,000		&14		&$48\pm12$		&$106\pm14$		&$-7.4\pm0.1$		&$-5.7\pm0.1$\\
2015.02.26		&VLT		&X-shooter	&1.0		&5\,400		&86		&$70\pm7$		&$143\pm7$		&$-7.1\pm0.1$		&$-5.5\pm0.1$\\
2015.11.14		&Keck		&HIRES		&1.1		&36\,000		&34		&$81\pm6$		&$153\pm8$		&$-7.0\pm0.1$		&$-5.6\pm0.1$\\

\hline

\end{tabular}
\end{center}

\flushleft
$^{\rm a}$ The listed uncertainties are rounded to the nearest 0.1\,dex and include only the model fitting errors (see 
Section 2.2).

$^{\rm b}$ This exposure was completed during twilight and suffers from high background emission.

\end{table*}

\begin{figure}
\includegraphics[width=84mm]{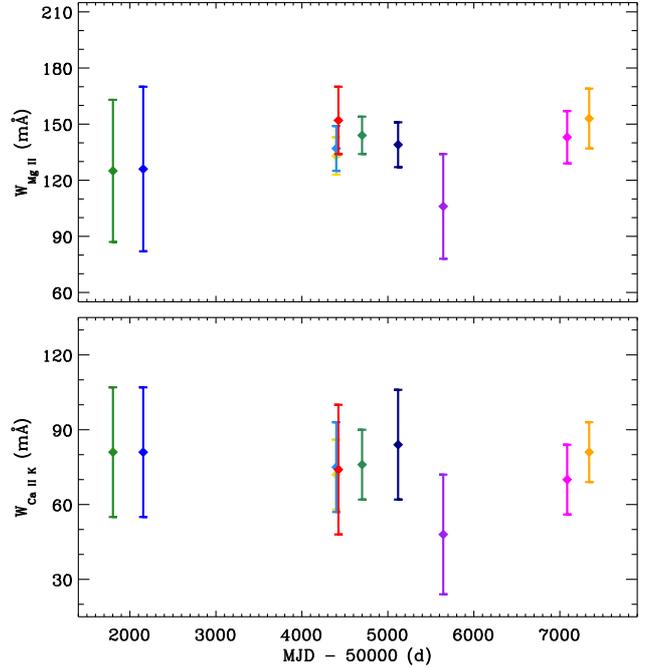}
\caption{The measured equivalent widths of the two metal absorption features at each epoch, color-coded identically 
to the spectra in Figure \ref{fig4}.  These measurements are shown with $2\upsigma$ error bars and indicate that no 
appreciable changes are observed in metal abundance, and are thus consistent with a constant accretion rate.  The 
purple data points corresponding to the 2011 Mar 23 data may have suffered from high background contamination and
problematic sky removal.
\label{fig5}}
\end{figure}

Equivalent widths ($W_\uplambda$) for Ca\,{\sc ii} K 3934\,\AA \ and Mg\,{\sc ii} 4481\,\AA \ are calculated for each 
epoch of spectroscopy and given in Table \ref{tbl2} and plotted in Figure \ref{fig5}.  It is noteworthy that the Ca\,{\sc ii} H 
3968\,\AA \ line is not detected at a significant level in any of the spectra, however, this transition is contained within the 
strong and broad H$\upepsilon$ absorption feature.  The spectral regions surrounding the lines are first normalized 
using a low-order polynomial, and the wavelength range used for the $W_\uplambda$ integration is shown in grey in 
Figure \ref{fig4}.  Uncertainties for the $W_\uplambda$ measurements are a combination of normalized flux errors and 
an estimate of the systematic uncertainty due to extraction and normalization of the spectra.  Normalized flux uncertainties 
are calculated by taking the standard deviation of 2\,\AA \ segments on either side of the line of interest.  The systematic 
uncertainty is set equal to the normalized flux uncertainty, and is summed in quadrature with the normalized flux errors to 
produce the total $W_\uplambda$ error (effectively increasing the total uncertainty by $\sqrt{2}$).



All spectra were fitted using white dwarf atmospheric models \citep{koe10} to determine abundances for calcium and 
magnesium.  The stellar parameters of GD\,56 were fixed at $T_{\rm eff} = 15\,000$\,K and $\log\,g=8.0$; this temperature
is within 1\% of the average literature value \citep{koe05,koe09,gia11}, and the surface gravity agrees well with the {\em
Gaia} DR2 parallax ($\varpi=14.07\pm0.05$\,mas; \citealt{gai18}) distance of $71.07\pm0.25$\,pc.  The fitting process 
worked well for all spectra with the exception of  the 2011 HIRES data set, where both the Mg\,{\sc ii} partially-resolved 
triplet and Ca\,{\sc ii} K line features are too narrow.  This suggests that these data are problematic owing to the high sky 
background, and they are disregarded in the subsequent analyses.  The uncertainties in metal abundance are only formal 
fitting errors and do not include other potential sources of error such as systematic offsets between instruments and their 
processing pipelines \citep{koe09b}.  In fact, the abundance values determined using the nominal instrument resolution 
reveal a clear instrument dependence, with offsets up to 0.4\,dex for [Ca/H] and up to 0.2\,dex in [Mg/H].  In order to better 
understand this issue, the resolving power for the UVES and HIRES model fits was increased up to 40\,000, and this was 
found not only to improve the agreement between all the abundance determinations, but also to better track the equivalent 
width measurements of both lines; these abundances are adopted here and listed in listed in Table \ref{tbl2}.  If the metal 
abundances in GD\,56 are constant, then the mean values and their dispersions of the reliable data sets are [Ca/H] 
$=-7.1\pm0.1$, and [Mg/H] $= -5.6\pm0.1$.

\section{DISCUSSION}

The changes in absolute infrared flux of GD\,56 are clear and compelling, but at the same time the flux ratios 
between the two shortest wavelength channels in both {\em Spitzer} and {\em WISE} are consistent with remaining 
constant (Figures \ref{fig1}--\ref{fig3}).  Concurrently, there is no indication that the photospheric line strengths or metal 
abundances are changing, although the sampling is relatively sparse.  Based on diffusion theory, the expected sinking 
timescales for calcium and magnesium in a hydrogen-atmosphere white dwarf with $T_{\rm eff}=15\,000$\,K, $\log\,g
=8.0$ are 1.0 and 0.9\,d respectively \citep{koe09}.  Thus, while the stellar photosphere should track any accretion rate 
changes on daily timescales or longer, none are apparent in the observations.  These results are comparable to those 
obtained via multi-epoch spectral observations of the dusty white dwarf prototype G29-38 \citep{von07,deb08}.

These observations provide empirical constraints on the possible underlying processes in the circumstellar disk 
orbiting GD\,56.  In this section, the infrared variability and apparently constant accretion rate are discussed in the 
context of relevant timescales, geometry, and possible physical models.  These ideas are considered in the context 
of other dusty systems for consistency, and with respect to disk evolution models to assess their wider applicability.  
The inferred, ongoing accretion rate is relevant for the following discussion, where diffusion theory predicts rates of 
$1.0\times10^8$\,g\,s$^{-1}$ and $0.6\times10^8$\,g\,s$^{-1}$ based on the magnesium and calcium abundances, 
respectively, assuming they are deposited at their bulk Earth mass  fractions \citep{all01,koe09}.

\subsection{Previous Infrared Modeling of GD\,56}

\begin{figure}
\includegraphics[width=84mm]{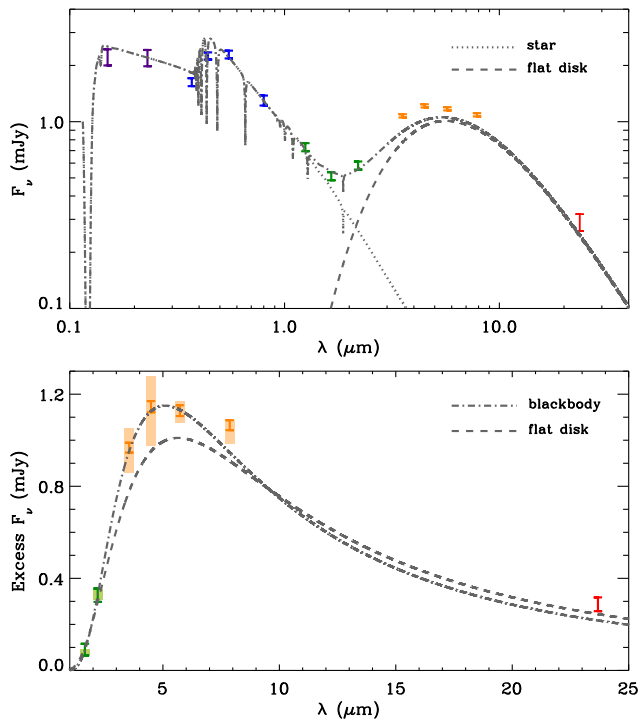}
\caption{The upper panel plots the photometric spectral energy distribution (SED) of GD\,56 from the ultraviolet through 
the infrared, including the first epoch of (cryogenic) IRAC data, and MIPS 24\,$\upmu$m flux.  The short wavelength data 
are fitted with a stellar atmosphere model of 15\,000\,K (dotted line) and added to this is a flat disk with inner temperature 
1400\,K, outer temperature of 600\,K, and zero inclination.  This flat disk model is representative of the best possible fit to 
the data with this restricted geometry, yet cannot account for the IRAC fluxes.  In the lower panel the stellar photosphere 
has been subtracted from the photometry, and this excess flux is plotted linearly with the same flat disk model, as well as 
a blackbody with $T=1000$\,K.  The shaded regions represent the range of observed infrared fluxes over 11.2\,yr, 
where the minima at 3.6 and 4.5\,$\upmu$m are conceivably (just) consistent with a face-on, flat disk.
\label{fig6}}
\end{figure}

It is worthwhile to first review previous data and modeling of the infrared emission in the GD\,56 system, as it remains 
unusual and to date has the largest, observed, fractional infrared luminosity of any dusty white dwarf \citep{roc15}.  Over 
a decade ago when GD\,56 was first observed with {\em Spitzer}, it was noted that the infrared emission could not be 
reproduced by a flat disk alone \citep{jur07}.  In Figure \ref{fig6} these initial infrared observations are reproduced with
better short wavelength data and atmospheric modeling, together with a flat disk model that fails to reproduce the IRAC
flux measurements.  Because it is now clear that the $3-5\,\upmu$m fluxes vary over time, any disk model must be viewed 
with caution when applied to infrared data that are not taken simultaneously, such as those in the figure.  Regardless, no 
flat disk model can reach the plotted fluxes, and thus this model is insufficient.  A modestly warped segment was later 
added to the flat disk model for GD\,56, and this was able to account for the overall infrared emission including its 
strong silicate feature \citep{jur09}.  The source of any warping in an otherwise flat disk is unclear, but possibilities 
include gravitational perturbations by orbiting bodies, radiative instability \citep{jur07}, and dust that follows the scale
height of gas.

In the lower panel of Figure \ref{fig6} the stellar photosphere has been subtracted and only the face-on (flat) disk model 
remains with the initial IRAC data, the sole MIPS 24\,$\upmu$m observation, and the $HK$ photometry.  The shaded 
regions give the extent of the observed variability to date in all bands with multiple measurements, where the largest 
variation is the 0.26\,mJy peak-to-peak flux change at 4.5\,$\upmu$m (between 2013 Oct and 2017 Dec).  While 
the temporal coverage in the infrared has (multi) year-long gaps, the overall behavior of the fluxes in Figure \ref{fig1} 
suggests the variable dust emission is either stochastic, or possibly a periodic brightening followed by dimming.  While 
radiative warping may be responsible for additional emitting surface that exceeds the intrinsic infrared brightness of a 
flat disk configuration, it does not necessarily follow that such a structure can cause the observed variability.

\subsection{Timescale and Accretion Rate Considerations}

The simplest model for the observed long-term infrared variability towards GD\,56 is the production and depletion 
of dust clouds that provide emitting surface areas in excess of -- {\em or in lieu of} -- a flat disk configuration.  Such 
clouds must not change orbital radius drastically, as this would change the dust temperature and hence the flux ratio 
between 3.5 and 4.5\,$\upmu$m.  A priori, any transient or periodic dust structure can be optically thick or optically 
thin, but it is noteworthy that blackbody dust with $T=1000$\,K \citep{jur07} will orbit near 1.46\,$R_{\odot}$ and thus 
somewhat outside the stellar Roche limit (for GD\,56, asteroid-like bodies with density 3\,g\,cm$^{-3}$ should disrupt 
only within 0.9\,$R_{\odot}$).  Before exploring possible mechanisms for dust production and depletion around GD\,56 
and other dusty white dwarfs, it is best to establish some theoretical context.

A good starting point is to consider the relevant physical timescales for particulate and gaseous debris in the vicinity 
of the Roche limit for white dwarfs, where the orbital radii are $r\approx 1.0-1.5\,R_{\odot}$.  The timescales that 
govern five processes are important: Keplerian orbits, condensation, collisions, Poynting-Robertson (PR) drag, and 
viscous spreading.  The orbital period at these radii for a $0.6\,M_{\odot}$ star is between 3 and 7\,h, and hence solids 
will orbit at (circular) speeds exceeding 250\,km\,s$^{-1}$.  Condensation of metallic gas onto grains is likely to be 
efficient in the case of identical chemical compositions, and take place on orbital timescales or less \citep{met12}.  
The collisional timescale for dust in a disk-like configuration is directly comparable to the orbital period  (i.e.\ $t_{\rm 
coll} \sim P/4\uppi\uptau$), with the optical depth $\uptau \gtrsim1$ in the case of a vertically opaque disk, and 
$\uptau \simeq L_{\rm IR}/L_*$ if the disk is optically thin in the radial direction.  A disk may be optically thick in the 
radial direction (due to grazing starlight), yet remain vertically optically thin and experience less frequent collisions 
than implied by $f = L_{\rm IR}/L_*$.  For white dwarfs with well-determined fractional dust luminosities \citep{roc15},
if their vertical optical depths are comparable to $f$, then collision timescales will be on the order of $1-20$\,d.  This 
is at least two orders of magnitude shorter than the timescale for angular momentum loss by solids due to PR drag,
where for 1\,$\upmu$m size grains with density 3\,g\,cm$^{-3}$, the timescales for GD\,56 are between 6 and 14\,yr.  
For the population of known dusty white dwarfs, all with $L_{\rm IR}/L_* \gtrsim 10^{-3}$, it can be shown that for 
$\uptau > 10^{-5}$, the collisional timescale is {\em always shorter than PR drag} \citep{far08}.

Before discussing the timescale for the viscous spreading of gas, the implications of collisions are considered.  In 
the restricted case of disks that are completely optically thin, the relevant timescale for solids is thus the collisional 
timescale.  For GD\,56 this is a few orbital periods and less than 8\,h within 1.5\,$R_{\odot}$, requiring that the entire 
disk be replenished every few days or so, and with a dust mass production rate sufficient to maintain $L_{\rm IR} /L_*
\approx0.03$ \citep{roc15}.  This entails around 10$^{18}$\,g \citep{jur09} per collision timescale, which would 
eventually result in a mass accretion rate onto the star that is higher than 10$^{12}$\,g\,s$^{-1}$, and consistent with 
numerical simulations of collisional cascades around white dwarfs \citep{ken17a}.  Such high rates of (sustained or 
transient) disk accretion have been suggested by theory and hinted at via observations \citep{raf11b,gir12,ken17b}, 
but so far have not been confirmed by (limited) X-ray data \citep{far18}.  Furthermore, the maximum accretion rate 
from PR drag alone acting on optically thin debris is set only by the stellar luminosity and the fraction of starlight 
intercepted by the disk \citep{boc11}; if all radiation from GD\,56 is intercepted by dust, the maximum mass accretion 
rate is $3\times10^{10}$\,g\,s$^{-1}$.  Thus if disks are primarily optically thin, and accretion rates do exceed this value, 
then collisions dominate the production of gas that eventually accretes and pollutes the star, but material must be 
continually supplied to the disk \citep{ken17a,ken17b}.

In the context of GD\,56, these potential disk implications can be compared with the stellar atmospheric data.  Based 
on the photospheric calcium and magnesium abundances and diffusion theory \citep{koe09}, the current total stellar 
accretion rate is $(0.8\pm0.2)\times10^8$\,g\,s$^{-1}$ using either of these two elements at its bulk Earth mass fraction 
\citep{all01}.  These estimates are orders of magnitude smaller than any of the above predictions for optically thin dust 
shell depletion -- especially where collisions may dominate the production of gas -- but agree well with model predictions 
for flat and opaque dust disks depleting via PR drag (i.e. \ no collisions \citep{raf11a,boc11}).  Notably, this accretion 
rate is typical for polluted stars with metal diffusion timescales of less than a few years, where an ongoing steady state 
can be confidently inferred -- based on diffusion theory, no system to date has been determined to substantially exceed 
$10^9$\,g\,s$^{-1}$ \citep{far16b}.  If correct, then any long-term ($10^4-10^7$\,yr; \citealt{gir12}) infrared emission in 
dusty white dwarf systems is likely to be from solids in a flat disk, and consistent with dynamical relaxation on orbital 
timescales, where a sufficient component is optically thick.  This represents the basic utility of the canonical flat and 
opaque disk model; immunity to collisional annihilation, longevity, and accretion rates that broadly match those of 
diffusion theory \citep{jur03,raf11a,met12}.  [It is noteworthy that recent simulations by \citet{ken17a,ken17b} cannot 
produce a vertically thin disk of material at relevant orbital distances for dusty white dwarfs, whereas analytical work 
\citep{boc11,met12} appears to favor this configuration.]

While collisions can be avoided by efficient damping in a flat and opaque disk \citep{far08,met12}, once solids have 
been sublimated or collisionally vaporized, regardless of geometry, the material will no longer evolve via radiation forces.
Prior to the accretion of material onto the stellar surface, the final relevant timescale may be set by gas viscosity.  And 
while the properties of pure gas disks orbiting white dwarfs have been discussed at length in the literature \citep{jur08,
raf11a,far12b,met12,ken17b}, the implications for infrared and optical variability have not been fully explored.  Several 
estimates exist for gas viscosity in a disk dominated by metals, where the two uncertain factors are in the size of the 
$\upalpha$ parameter \citep{sha73} and whether the disk is ionized or neutral.  For bulk Earth material, the mean atomic 
weight of neutral gas would be 32.8\,$m_{\rm p}$, and if this material is fully ionized then 16.4\,$m_{\rm p}$.  Using these 
two extrema yields viscous timescales in the range $(10/\upalpha$)\,yr $\la t_{\upnu}\la(20/\upalpha$)\,yr for a typical white 
dwarf disk at $r=0.2\,R_{\odot}$.  Only in the limit $\upalpha\approx1$ can the viscous timescale approach that of PR 
drag, and only then for grains larger than 1\,$\upmu$m \citep{raf11a}.  

Therefore, a significant time lag may result between a change in the rate of solids delivered into an inner gas disk, 
and a subsequent change in the mass accretion rate from this reservoir.  If $\upalpha\approx0.1-0.4$ as inferred for 
a variety of fully ionized accretion disks \citep{kin07}, then the shortest realistic timescale would be $t_{\upnu}\approx
25-100$\,yr.  Therefore any changes in dust and gas production within a disk, including the inward movement of solids 
that eventually sublimate, can be mediated prior to accretion onto the stellar surface if an $\upalpha$-type gas disk 
is present.  Any such inner disk will smooth over changes on these relatively long timescales, and no changes in 
accretion rate would be expected on PR drag timescales.  While the presence of such inner gas disks orbiting white 
dwarfs is empirically unconstrained, if they are ubiquitous then changes in accretion rate might not be expected on 
human timescales.  There are a handful of circumstellar gas detections via absorption towards polluted stars; most
dramatically around WD\,1145 in the optical \citep{xu16,red17}, and possibly via a weak circumstellar component in 
the Ca\,{\sc ii}\,H and K lines of EC\,11246--2923 \citep{deb12a}, but also in the far-ultraviolet spectra of both 
SDSS\,1228+1040 and PG\,0853+516 \citep{gan12}.  In none of these cases do the gas velocities approach that
expected for free-fall, and hence the behavior of disk material prior to reaching the stellar surface is unknown.  An 
$\upalpha$ disk reservoir of gas is consistent with the unchanging metal lines in GD\,56 and SDSS\,0959, despite 
the changes observed in infrared emission \citep{xu14}, and regardless of the exact dust production or depletion 
mechanisms.  

Because this study considers non-canonical disk geometry and evolution for GD\,56, it is worthwhile to mention 
the one case where a non-canonical configuration has been established.  The {\em circumbinary disk} orbiting the 
polluted white dwarf SDSS\,J155720.77+091624.6 (hereafter, SDSS\,1557) and its companion must be optically thin 
and vertically extended to account for the infrared data \citep{far17}.  The system exhibits a strong infrared excess 
that cannot be reproduced by emission from a cool companion, and prior to the discovery of its duplicity, the infrared 
emission was well-matched by a face-on, flat disk model \citep{far12a}.  However, the orbiting companion dynamically 
precludes material in the region where such a flat disk would reside, and thus the $T\approx1100$\,K dust emission 
must originate from optically thin material at $r=3.3\,R_{\odot}$ \citep{far17}.  SDSS\,1557 demonstrates the plausibility 
of a non-flat disk of material that continuously feeds atmospheric metals onto a white dwarf for at least several years.

\subsection{Infrared Brightening and Dimming at GD\,56}

A plausible scenario for infrared variability in white dwarf dust disks is the following.  There are clouds of dust 
that are transiently or quasi-periodically appearing and disappearing in a circumstellar environment with an 
underlying disk that is substantially more massive.  Such a picture can account for the rapid decline of infrared 
flux from SDSS\,0959, where the discovery observations were taken during the presence of additional clouds of 
material, and these were quickly re-assimilated into an existing disk.  Any debris cloud generated by impact or 
collisions will have a range of orbits, {\em but all will intersect that of any underlying disk}, unless it occurs at
the extremities.  For a stray body, an impact onto the innermost edge of a particulate disk within the Roche limit 
is the least probable location, and requires a body or bodies smaller than a km that were immune to tidal disruption 
on any prior periastra \citep{bro17}.  Unless optically thick, any transient dust cloud that is vertically offset from an 
extant disk plane would quickly attain a higher temperature than debris that is partly shielded from starlight.

While transient dust clouds quickly being absorbed into an existing disk may be able to account for the sparse 
infrared data on SDSS\,0959, it may not work for GD\,56 where the infrared dimming takes place over many years.  
It is possible, however, that a simple (re-)incorporation of material above or below the plane of an underlying disk 
may be a process that acts like a damped oscillator.  If so, then dust clouds with relatively small masses may 
quickly become assimilated into a larger disk if damping is near critical, but the system may oscillate if damping 
is inefficient or the dust cloud mass is larger.  In the canonical opaque disk, damping is sufficiently high that 
collisions are negligible, and (re-)condensation of gas onto a particulate disk should occur on orbital timescales
\citep{met12}.  This scenario might account for the disappearance of gas emission lines around 
SDSS\,J161717.04$+$162022.4 \citep{wil14}.  

In the case of GD\,56, an important observational clue to the dust removal process is the fact that the flux ratios 
between 4.5\,$\upmu$m and 3.5\,$\upmu$m do not change substantially.  This implies the dust is created, and 
later destroyed or subsumed, within a range of orbital radii that does not change substantially with time.  Another 
interesting empirical indication is that the gradual decay in $3-5\,\upmu$m flux around GD\,56 appears to mimic 
that seen towards several bright debris disks orbiting young main-sequence stars, and where this decay is thought 
to be due to collisions of mm-size condensates \citep{men14,men15}.  In these studies, the generation of extreme 
infrared excesses are attributed to planetary impacts.  These impacts produce vaporized debris, which re-condenses 
and undergoes a collisional cascade that eventually depletes grains that emit efficiently in the infrared, aided by the 
removal of the smallest grains by radiation pressure.  While the blowout of dust is not relevant for white dwarf systems, 
the fact that the collisions will dominate unless damped suggests the analogy here may be compelling.  Removal of 
small grains around white dwarfs might occur via continued collisions and sputtering, but alternatively if they are 
heated to temperatures sufficient for rapid sublimation as may be the case for WD\,1145 \citep{xu18}, then there 
could be sufficient means to destroy dust in the vicinity of where it is produced.  It is noteworthy that the channel 
2:1 flux ratios do not appear to change substantially between the combined IRAC and {\em WISE} epochs reported 
for SDSS\,0959 \citep{xu14}.

In further analogy with WD\,1145, it is plausible that minor bodies are generating clouds of $T\approx1000$\,K
dust around 1.46\,$R_{\odot}$ either via collisions or sublimation -- but not tidal disintegration at this distance.  
It is now well established that the disk of dust and gas orbiting WD\,1145 is eccentric \citep{cau18}, and the 
same appears to be the case for at least two other disks \citep{man16a,den18,mir18}.  This implies that tidal 
disintegration of a large body or bodies is not likely to be ongoing in a near-circular orbit at or near the Roche 
limit \citep{gur17,ver17}, but instead orbiting bodies are crossing in and out of this radius.  Collisions associated 
with circularization of a disk may produce dust periodically, some of which may be located outside of the disk 
plane, and later this material becomes assimilated into the evolving disk.  There are a few considerations that 
make this scenario plausible, as long as the dust production sites are relatively localized and not varying 
in orbital distance on the timescales probed by the observations.

First, the circularization process is not well constrained, and material may end up over a range of orbits 
spanning the Roche limit.  This could imply that intact bodies survive longer outside the Roche limit, and are 
thus available as sites of dust production.  Second, the Roche limit separates disrupted and intact bodies of 
the same composition, but there may be a transition zone where some bodies remain intact while others are 
fragmented due to compositional gradients.  Third, small bodies can be formed by recycled ring material at
the Roche limit via disk spreading \citep{van18} -- as has been suggested for the rings of Saturn \citep{cha10}
-- and this process might also play a role in the rise and fall of emitting dust.  This latter scenario requires a 
massive disk in excess of a lunar mass, but at the same time GD\,56 has the largest known infrared excess 
of any dusty white dwarf.

While somewhat speculative, the infrared data for GD\,56 may hint at an overall trend that includes a flux 
increase followed by a gradual decay.  If there is a periodic signal here, it would be crudely around 7\,yr and 
then correspond roughly to an orbital scale of 3\,AU.  In this case an infrared brightening might be expected
before or near 2020.  It is not unreasonable that some tidally-disrupted material remains in the process 
of circularization, with small bodies still strung out along a range of wide orbits \citep{deb12b,ver14}, and these 
periodically interact with the disk around GD\,56, generating dust that is eventually destroyed or re-incorporated.  
Another remote possibility is if a planetary body on a wider orbit has sufficient gravitational influence to induce 
a disk warp during periastron, it could also be that a warp may decay gradually as a damped oscillator (but in 
the case of GD\,56 it would need to be highly inefficient with a decay timescale longer than $10^3$\,d).  White
dwarf disk precession via general relativity can be on the order of years, with a steep dependence on radial 
distance \citep{mir18}, but it is unclear if this would drive a periodic warp or change in dust mass.  More infrared 
data with better temporal coverage is required before any realistic model constraints can be made.

Lastly and importantly, the dimming trend at GD\,56 after MJD 56\,000 appears to be consistent with the broad 
behavior predicted by numerical simulations of collisional cascades around white dwarfs \citep{ken17a}, and 
more generally the predictions of steady state evolution of debris disks due to collisions \citep{wya07}.  In the
former simulations, the basic behavior of the infrared dust luminosity is stochastic, yet switching between high 
states of visibility, and non-detectable low states.  Steady states are possible, or even likely with continued 
mass input into an existing disk, but without such input an infrared excess will decay to non-detectable levels.  
Nevertheless, the decay trend seen in the data for GD\,56 is well within the family of predictions based on
these models, and while solutions are degenerate, roughly speaking the model sets that can reproduce the
infrared data have dust and gas production rates in excess of $10^{12}$\,g\,s$^{-1}$ and total belt masses in 
excess of $10^{23}$\,g \citep{ken17a}.

While such rates are consistent with the estimate made above based on destroying all the infrared emitting 
dust around GD\,56 within a typical collision timescale, they are far in excess of that inferred to be ongoing 
via diffusion theory.  Such high accretion rates, if they occur, have fundamental implications for the nature of 
the remnant planetary systems orbiting white dwarfs.  For example, if 30\% of white dwarfs with cooling ages 
in the range $20-200$\,Myr accrete from disks \citep{koe14} that produce accretion rates as high as $10^
{13}$\,g\,s$^{-1}$, then a given system with a duty cycle of $0.3 \times 100$\,Myr can have accreted up to 
$10^{28}$\,g (2\,$M_{\oplus}$) of material.  This is in stark contrast to the consumption of a mass closer to 
$10^{25}$\,g (Pluto) based on the more sedate accretion rates inferences made from diffusion theory 
\citep{koe09}.  This potential tension between theory and observation cannot be resolved within the context 
of the current work, but the infrared behavior of GD\,56 is a strong indication that collisions are of fundamental 
importance in the debris disks that orbit polluted white dwarfs.

\section{Summary and Current Perspective}

Infrared observations of GD\,56 reveal both brightening and dimming events associated with circumstellar dust.  
Over a span of 11.2\,yr, the $3-5$\,$\upmu$m fluxes in both {\em Spitzer} and {\em WISE} are shown to increase 
and decrease, where the peak-to-peak changes are over 20\% and a gradual dimming is apparent for at least one 
and possibly two intervals.  In contrast, neither the flux ratios between the two shortest wavelength channels on 
both spacecraft, nor the photospheric metal absorption in the star appear to be changing.  These collective results
suggest dust is produced and later destroyed (or subsumed) without any significant change in the range of orbital
radii; i.e.\ PR drag is not affecting the distribution of the emitting dust.  Any gas resulting from the dust production 
process -- including small dust grains that rapidly sublimate -- may quickly re-condense onto grains or join an 
$\upalpha$-like accretion disk, and this may delay changes in mass accretion rate for decades or longer.  
These ideas have applicability beyond GD\,56 and are generally consistent with all observations to date
for polluted white dwarfs from the infrared to the optical.

It is hypothesized that collisions with an existing and evolving disk may lead to infrared brightening via dust 
production, and, subsequently, infrared dimming as the cascade grinds down the material into sizes that do not
emit efficiently at micron wavelengths.  Competing effects may be present if there are massive, flat and opaque 
disks orbiting most polluted white dwarfs, as transient debris clouds will be readily subsumed into such an
underlying disk.  A canonical disk may be warped by gravitational forces, and in this way increase its emitting 
surface to produce infrared flux changes, but it would likely require a fairly massive orbiting body, and the
observed dimming may be difficult to reconcile with such a scenario.  Of all the ideas discussed here, only 
variable disk warping will not give rise to an eventual change in accretion rate onto the star (and in the
broader context cannot account for the continued presence of gas in many debris disks orbiting polluted
stars).

On the one hand, the idea that white dwarf disks are undergoing a collisional cascade is rather reasonable, 
and for GD\,56 the infrared flux changes have clear similarities with those observed towards young, planet-forming
debris disks orbiting main-sequence stars.  This can also account for the continued production of gas for many stars
that exhibit these features by emission or absorption or both.  On the other hand, and despite broad agreement with 
model families of collisional cascades, the total disk masses and production rates for dust and gas are orders of 
magnitude higher than inferred from diffusion theory, and the disk never settles dynamically into a flat configuration.
Recently, \citet{bau18}, building on previous work \citep{dea13,wac17}, find ongoing accretion rates up to $10^{13}
$\,g\,s$^{-1}$ are possible if hydrogen-rich white dwarfs suffer the thermohaline instability in their mixing layers 
where metals are present.  While this possibility has been discounted in the context of white dwarf atmospheres
and diffusion theory \citep{koe15}, it has clear parallels with the dust production rates via collisions discussed in
this work inspired by GD\,56.  While diffusion theory predicts that any {\em changes} in mass accretion rates should 
be observable as {\em changes} in their metal absorption features over a finite number of sinking timescales, the
models that include thermohaline mixing do not \citep{bau18}.  Thus, on the one hand, accretion rate changes may 
be expected from diffusion theory, yet restrained by longer gas viscosity timescales.  But on the other hand, even if
accretion occurs directly onto the stellar surface, changes in metal abundance would not be observable if a 
thermohaline instability stifles diffusion.

It is ironic that this discovery comes 15 years after the launch of {\em Spitzer}, and towards the end of its extended, 
post-cryogenic lifetime.  \citet{rea09} noted that IRAC short-cadence monitoring of G29-38 -- the prototype and 
brightest dusty white dwarf -- revealed $3-8$\,$\upmu$m flux changes, but as a well-known pulsating variable 
of the ZZ\,Ceti class such changes are particularly challenging to interpret as variability in the circumstellar dust 
content.  While likely nearing the end of their mission lifetimes, both {\em Spitzer} and NEOWISE currently offer 
the best observational constraints on the evolving and dusty planetary systems orbiting polluted white dwarfs.

\section*{Acknowledgements}

The authors acknowledge useful conversations with R. Rafikov, and thank an anonymous reviewer for a careful
reading of the manuscript.  This work is based in part on observations made with the {\em Spitzer Space Telescope}, 
which is operated by the Jet Propulsion Laboratory, California Institute of Technology under a contract with NASA.
This publication makes use of data products from the {\em Wide-field Infrared Survey Explorer}, which is a joint project 
of the University of California, Los Angeles, and the Jet Propulsion Laboratory / California Institute of Technology, funded 
by the NASA.  Some of the data presented herein were obtained at the W. M.\ Keck Observatory from telescope time 
allocated to NASA through scientific partnership with the California Institute of Technology and the University of California.  
This work was supported by a NASA Keck PI Data Award, administered by the NASA Exoplanet Science Institute.  Some 
of the observations presented here were made with ESO Telescopes at the Paranal Observatory.  JF acknowledges 
support from STFC grant ST/R000476/1.  RvL was supported by the DISCSIM project, grant agreement 341137 funded 
by the European Research Council under ERC-2013-ADG.  TGW wishes to acknowledge funding from a STFC studentship, 
and OT was partially supported by a Leverhulme Trust Research Project Grant.  Research leading to these results has 
received funding from the ERC under the European Union's 7$^{\rm th}$ Framework Programme no.\ 320964 (WDTracer).  
TvH was supported by the National Science Foundation Award AST-1715718, and AB acknowledges the support of a 
Royal Society Dorothy Hodgkin Fellowship.

\bsp    
\label{lastpage}
\end{document}